\RequirePackage{amsmath}
\documentclass[runningheads]{llncs}
\usepackage{graphicx}
\graphicspath{{pics/}} 

\usepackage{listings}
\usepackage{xcolor}
\colorlet{punct}{red!60!black}
\definecolor{background}{HTML}{EEEEEE}
\definecolor{delim}{RGB}{20,105,176}
\colorlet{numb}{magenta!60!black}
\lstdefinelanguage{json}{
    basicstyle=\scriptsize\ttfamily,
    numbers=left,
    numberstyle=\scriptsize,
    stepnumber=1,
    numbersep=8pt,
    showstringspaces=false,
    breaklines=true,
    frame=lines,
    backgroundcolor=\color{background},
    literate=
     *{0}{{{\color{numb}0}}}{1}
      {1}{{{\color{numb}1}}}{1}
      {2}{{{\color{numb}2}}}{1}
      {3}{{{\color{numb}3}}}{1}
      {4}{{{\color{numb}4}}}{1}
      {5}{{{\color{numb}5}}}{1}
      {6}{{{\color{numb}6}}}{1}
      {7}{{{\color{numb}7}}}{1}
      {8}{{{\color{numb}8}}}{1}
      {9}{{{\color{numb}9}}}{1}
      {:}{{{\color{punct}{:}}}}{1}
      {,}{{{\color{punct}{,}}}}{1}
      {\{}{{{\color{delim}{\{}}}}{1}
      {\}}{{{\color{delim}{\}}}}}{1}
      {[}{{{\color{delim}{[}}}}{1}
      {]}{{{\color{delim}{]}}}}{1},
}

\usepackage{tabularx}
\usepackage{booktabs}

\usepackage{amsthm}
\usepackage{thmtools}
\declaretheoremstyle[
spaceabove=6pt, spacebelow=6pt,
headfont=\normalfont\bfseries,
notefont=\mdseries, notebraces={(}{)},
bodyfont=\normalfont,
postheadspace=0.6em,
headpunct=:
]{mystyle}

\usepackage{cleveref}
\crefname{hyp}{}{}
\Crefname{hyp}{}{}

\usepackage[htt]{hyphenat}

\usepackage{subcaption}
\usepackage{url}

\usepackage{enumitem}

\begin{document}
\title{Towards Secure and Leak-Free Workflows\\ Using Microservice Isolation\thanks{This project has been made possible in part by a grant from the Cisco University Research Program Fund, an advised fund of Silicon Valley Foundation.}}
%
%
\author{Loïc Miller, Pascal Mérindol, Antoine Gallais and Cristel Pelsser}
%
\authorrunning{L. Miller et al.}
%
\institute{University of Strasbourg}

\maketitle              
\begin{abstract}
Data leaks and breaches are on the rise.
They result in huge losses of money for businesses like the movie industry, as well as a loss of user privacy for businesses dealing with user data like the pharmaceutical industry.
Preventing data exposures is challenging, because the causes for such events are various, ranging from hacking to misconfigured databases.
Alongside the surge in data exposures, the recent rise of microservices as a paradigm brings the need to not only secure traffic at the border of the network, but also internally, pressing the adoption of new security models such as zero-trust to secure business processes.

Business processes can be modeled as workflows, where the owner of the data at risk interacts with contractors to realize a sequence of tasks on this data.
In this paper, we show how those workflows can be enforced while preventing data exposure.
Following the principles of zero-trust, we develop an infrastructure using the isolation provided by a microservice architecture, to enforce owner policy.
We show that our infrastructure is resilient to the set of attacks considered in our security model. We implement a simple, yet realistic, workflow with our infrastructure in a publicly available proof of concept.
We then verify that the specified policy is correctly enforced by testing the deployment for policy violations, and estimate the overhead cost of authorization.

\end{abstract}
\section{Introduction}
\label{sec:introduction}

Data leaks and breaches are increasingly happening.
With more and more businesses using public clouds to process data~\cite{cloud-adoption}, and this data being frequently moved around, exposures are more likely to happen than ever.
Those exposures are perceived as huge losses of money for businesses like the movie industry~\cite{byers2003analysis}, and as a loss of user privacy for applications dealing with user data~\cite{privacy-rights}.
There has been a steady increase in the number of breaches reported over the past eight years~\cite{quickview-report}, 2019 being an all-time high with 5183 reported breaches as well as 7995 million records lost~\cite{quickview-report}.
Malicious actors have been responsible for most incidents, but accidental exposure of data on the Internet (e.g. misconfigured databases, backups, end points, services) has put the most records at risk~\cite{quickview-report}.

Even though both data breaches and data leaks result in data being exposed to unauthorized entities, the way this data is exposed is different.
On one hand, data breaches refer to the unauthorized access of data by exploiting flaws in the security of the breached system.
Data breaches can happen with data at rest~\cite{yahoo-breach} where attackers exploit a flaw to gain access to the data, or in transport~\cite{mikrotik-breach}, where attackers exploit a vulnerability to eavesdrop on traffic.
On the other hand, data leaks refer to the exposure of data belonging to an entity, due to the way this data is processed by this entity, by a mistake~\cite{fafc-leak}, or caused by malicious behavior~\cite{google-leak}.

Preventing both forms of data exposure is a challenging problem to tackle, since an exposure can occur in multiple circumstances, caused by hacking, misconfigured databases, malicious insiders, etc.~\cite{privacy-rights}.
As attacks were generally assumed to come from a location external to the system via north-south traffic, the traditional way to secure such a system against those attacks was to protect it at the border via gateways, firewalls or programmable switches~\cite{jin2016understanding}.
The recent rise of microservices as a paradigm, and their increased use in building large, cloud-based enterprise applications~\cite{chandramouli2020building} brings new security risks, increases the attack surface and makes protecting the border no longer sufficient.
To prevent data leaks, one also needs to consider attacks coming from inside the system (e.g. leaks stemming from the way data is processed or caused by a malicious employee).
Not only does north-south traffic need to be protected, but so does the traffic between services within the system. 
The zero-trust security model~\cite{gilman2017zero}, where all traffic flows are required to be authenticated and authorized via fine-grained policies, provides such protection.

In accordance with the principles of zero-trust, we want to achieve a zero-trust secure system that enables the exchange of data between non-trusted agents while guaranteeing this data is secure at rest, in transport and cannot be exposed by any agent in both cases.
More specifically, we want to guarantee those properties in the context of a workflow.
We define a workflow as a sequence of tasks performed by a set of independent actors.
The owner of the data (i.e., the instigator of the workflow) interacts with contractors to realize a sequence of tasks.
Both the owner and the contractor have agents processing the data, which can either represent an employee or a fully automatic service.


In the context of a workflow, security mechanisms quickly become challenging to configure, manage, scale and monitor when combined, with a large number of actors using different IT environments.
Additionally, keeping a monolithic approach can lead to switching cost and vendor lock-in issues.
A system designed to prevent data exposures must thus be simple, modular, scalable and easy to manage.
The microservice architecture fulfills those needs thanks to its loosely coupled services.

In recent years, microservices have become the de-facto standard way of developing cloud-native apps~\cite{chandramouli2020building}.
Among the different ways microservices can be deployed, service meshes have been characterized as the latest evolution in software service design, development and delivery~\cite{jamshidi2018microservices}.

To meet the additional requirements for zero-trust and prevent data leaks, we need specific control over the environments the agents will be using, to make sure that all the actions of an agent follow a policy enforced by the owner.
The environment in which the agents execute their task is isolated and secured thanks to traffic interception and access control policy enforcement.
In our approach, we choose to use the microservice architecture to do so.
In our infrastructure, agents of the workflow are mapped to containers, which are then used in conjunction with an orchestrator, a service mesh and policy engines to enforce the workflow as well as the policy of the owner.

Considering the problem at hand (Sec.~\ref{sec:problem}), we specify our desired properties, a threat model and then define a security model.
We describe our proposed solution (Sec.~\ref{sec:descr}), and detail how we can enforce our policy over the workflow.
We then describe the proof of concept we realized alongside our solution (Sec.~\ref{sec:poc}), and review how our infrastructure is resilient to the set of attacks considered in our security model as well as estimate the overhead cost of authorization (Sec.~\ref{sec:measures}).
Finally, we discuss the benefits of the infrastructure (Sec.~\ref{sec:discuss}), review related work (Sec.~\ref{sec:rel}) and conclude (Sec.~\ref{sec:conclusion}).

\section{Problem Statement}
\label{sec:problem}

In the context of a workflow, the system we want to achieve should enable the secure exchange of data and its security at rest while avoiding any leak.
We define a workflow as a sequence of tasks performed by a set of independent actors.
The owner of the data (i.e., the instigator of the workflow) interacts with contractors to realize a sequence of tasks.
The workflow is defined by the owner, which defines how and by whom the data he possesses should be processed and specifies the different steps needed to achieve his objective.
Those tasks are realized by contractors, which perform the task they have been assigned to, on the data they have been given.
Contractors possess some business intelligence, which we define as the tools and methods used to fulfill their task(s).

Both the owner and the contractor have \textbf{agents} processing the data, which can either represent an employee or a fully automatic trusted service.
Agents fulfill (part of) the task assigned to an actor.
The contractor fragments their task as they see fit between their agents.

The owner has ownership over the data being processed: he does not want his data to be leaked in any way.
On the other hand, the contractors do not want the other actors involved in the workflow, including the owner, to learn about their business intelligence. 

For example, let us consider the case of a workflow, where an owner in the post-production stage of making a movie wants to employ other companies to edit the video and audio components of the movie~\cite{byers2003analysis}.
More specifically, let us imagine that for example the owner wants to add special effects, tune colors, set up High Dynamic Range (HDR) and master the audio. In particular, he wants the application of the special effects first, and then the color tuning, and, finally, HDR in parallel with the sound mastering.

\begin{figure}[htbp]
 \centerline{\includegraphics[width=0.6\linewidth]{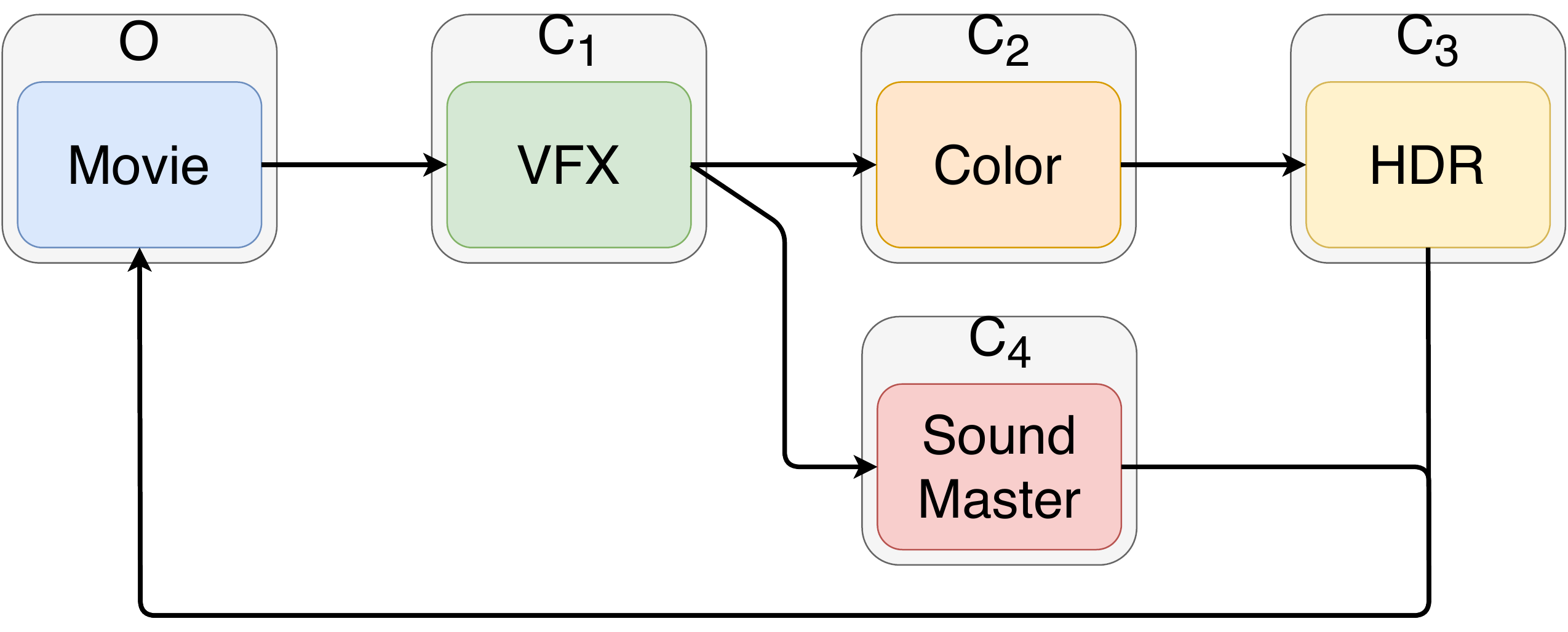}}
 \caption{Movie workflow example. Arrows model the specified workflow, and thus represent the communication flow of our example. The owner ($O$) sends its data to the first contractor ($C_1$), for special effects processing. $C_1$ then sends the modified data along the workflow, for color ($C_2$), HDR ($C_3$) and sound ($C_4$) processing. The resulting data is then sent back to the owner.}
 \label{fig:wfs}
\end{figure}

The intent of the owner can be modeled under the form of a workflow, that is a directed acyclic graph as depicted on Fig.~\ref{fig:wfs}.
The owner ($O$) will first send its data to the company responsible for special effects ($C_1$).
$C_1$ will then apply its business intelligence to add the special effects to the data the owner sent to him, and then send the result to the company responsible for coloring ($C_2$) as well as the company responsible for sound mastering ($C_3$).
$C_2$ will then send the result of its modifications to the company responsible for HDR ($C_3$) and sound mastering ($C_4$).
Finally, both $C_3$ and $C_4$ will send the result of their modifications back to the owner.

The purpose of our solution is to allow this workflow to take place, with the movie transiting from the owner ($O$) through the contractors ($C_1, C_2, C_3, C_4$) back to the owner, while guaranteeing data security at rest and in transport.
With our proposal, we prevent unwanted communications at the network and application level.

\subsection{Desired Properties \& Threat Model}

With our solution, we want to guarantee the following properties.
\begin{itemize}
 \item \textbf{Data security at rest}: Data within the workflow is stored encrypted (confidentiality) to prevent malicious entities from reading the data.
 Access to the data is restricted by isolation.
 This data, as well as the business intelligence of the contractors, cannot be leaked outside of the workflow, before, during or after the workflow is terminated. 
 \item \textbf{Data security in transport}: Data is exchanged encrypted (confidentiality) to prevent eavesdroppers from reading the data.
 The data is also transmitted accurately and complete (integrity), with a verified origin and destination (authentication).
 This data, as well as the business intelligence of the contractors, cannot be leaked outside of the workflow, before, during or after the workflow is terminated.  
\end{itemize}

We consider a threat model from the point of view of each actor of the workflow.
\begin{itemize}
 \item \textbf{Owner}: The owner does not want the data it sends to the contractors involved in the workflow to be leaked.
 The threat here is that an agent leaks the data of the owner to an unauthorized party, or that the data is accessed by an unauthorized adversary.
 \item \textbf{Contractor}: The contractors do not want the other actors involved in the workflow to learn about their business intelligence.
 The threat in this case is that actors learn about the processes used by a contractor.
		For example, actors might be able to guess the set of actions applied on the data or the algorithms used to transform the data.
\end{itemize}

\subsection{Trust Model: Actors and Environment}

From the point of view of the data owner, trusting the contractors is one thing, trusting its agents another.
In other words, if the owner trusts the organization of the contractor to not intently bypass our system, controlling the actions of the contractor's agents is then possible for both the owner and the contractors.
If one does not trust the contractors to deploy the infrastructure they are required to deploy, there is no easy way to verify that the data is actually sent to the secure environment we designed (Sec.~\ref{sec:descr}), therefore removing any guarantee we might have concerning data leaks.
Removing this trust between actors has its own drawbacks in a real world deployment (see App.~\ref{sec:trust}).

In contrast, looking at a finer granularity, actors do not need to trust their agents and the ones of the other actors.
Even though agents are deterred from engaging in malicious activities, due to the nature of their relationship with their companies (internal rules, non-disclosure agreements, ...), they can still put data and/or business intelligence at risk through accidental exposure or malicious behavior.
Actors are thus assumed to be malicious.
Our solution controls those agents to prevent owner data and business intelligence leaks.
This is consistent with our need to trust the contractors, since business to business contracts have the same deterrents, but with much higher stakes at play.

From the point of view of a contractor, we have the same trust issues as the owner.
Other actors, including the owner, might try to reverse engineer the business intelligence of the contractor.
This reverse engineering process requires a lot more effort than simply having access to the data of the owner, and might prove to be very hard or even impossible to do in some cases\footnote{\footnotesize{This can happen in very specific cases, such as when a contractor receives its input(s) and gives its output(s) to the same actor. As data is encrypted in transport, only the two ends of a communication see the data. A solution would be to insert the owner between contractors such as to limit their learning of the workflow and trust the owner not to reverse engineer the actions of its contractors}}.
In the same way the owner needs to trust that contractors do not intently bypass our system, the contractors need to trust that actors sending them data do not tamper with it.
Like the owner, contractors do not trust the agents.

Finally, both the owner and the contractors need to trust the owners of the environments involved in the workflow.
While the environment an actor is using can be owned by this actor, meaning the added trust requirement is the same as trusting the actor, some actors can use a third-party environment to fulfill their task(s) (e.g. a cloud provider).
Since this third-party provides (part of) the environment the workflow will be deployed on and has admin rights to the machines supporting the workloads, it can try to gain access to the data of the owners or the business intelligence of the contractors.
We would need to enhance our solution with Trusted Execution Environments (TEEs) in order to fully remove the need for trust in those third-parties (again, see App.~\ref{sec:trust} for the implications of removing this trust).
With the proposed infrastructure, one needs to trust those potential third-parties.
As such, actors and environment providers are considered honest but curious.

To summarize, from the point of view of the owner or a contractor, we trust everything but the agents.
Actors are assumed to be honest but curious, while agents are assumed to be malicious.

\subsection{Attacker Model: External Attackers and Malicious Agents}

Taking into account the assets to protect and the trust model, we consider three types of attackers in our model.
As our solution is in the form of a deployed infrastructure, an attacker can be internal or external to the modeled workflow.
An external attacker can then either be co-located with the deployed workflow (e.g. the attacker is located in a cloud partially or fully hosting the workflow) or external, attacking from a remote location.
\begin{itemize}
 \item \textbf{External attacker}: External to the workflow and the location of the infrastructure deployment.
 This highly motivated and often technically skilled attacker tries to gain access to the data or the business intelligence of the contractors from the outside.
 \item \textbf{Co-located attacker}: External to the workflow, but co-located at the deployment (e.g. an attacker located in one of the clouds that are part of the workflow deployment).
 This highly motivated and often technically skilled attacker tries to gain access to the data or business intelligence of the contractors, but from a co-located position that opens more exploit possibilities.
 \item \textbf{Malicious agent}: Internal to the workflow, this attacker tries to leak the data outside of the workflow.
\end{itemize}

Despite the fact our model already covers most cases, it does not deal with the full range of possible attacks.
Fully protecting against some attacks (e.g. from a contractor or a third-party cloud provider) would make the system less convenient and usable for contractors or the owner (see App.~\ref{sec:trust} for the implications).
Protection from leaks resulting from physical attacks such as when an employee takes a picture of his screen are not considered.

\section{Isolate Agents with a Microservice Model}
\label{sec:descr}

We now present the infrastructure we propose for protecting a workflow execution from the threats expressed in Sec.~\ref{sec:problem}.
As we need a way to prevent data leaks, we need to control the communications an agent can engage in.
To achieve this, we need to control the environments the agents will be using, to make sure that all the actions of an agent follow a policy enforced by the owner.
We opted to do this using the microservice architecture, for the benefits granted by the components of the infrastructure listed below.
Besides, they are already commonly deployed to provide many services like automatic scaling and isolation.

\subsection{Overall Description of the Infrastructure}

In this infrastructure, agents of our workflow are mapped to containers, which are then used in conjunction with an orchestrator, a service mesh and policy engines to enforce the policy of the owner:
\begin{itemize}
 \item \textbf{Container}: Standard unit of software packaging application code, called a service, and its dependencies in an isolated environment.
 Containers grant us portability and a standardized environment that we believe can be easily adopted by the contractors.
 Agents of the workflow are each mapped to a container.
 Some examples of container technology include Docker~\cite{docker} and Linux containers (LXC, LXD)~\cite{linux-containers}
 \item \textbf{Orchestrator}: An orchestrator is a system used to automate the management of containers and their life cycles.
 The orchestrator does this by interacting with workers running on physical machines, which control groups of containers called pods.
 Some examples of orchestrators include Kubernetes~\cite{kubernetes} and Nomad~\cite{nomad}.
 \item \textbf{Service Mesh}: A service mesh is a system used to automate the communication, security and monitoring of containerized environments.
 The service mesh controller configures proxies for each environment, which are added as a sidecar to the pod or embedded directly into the services.
 Those proxies act as intermediaries between agents to secure and control their interactions.
 In other words, proxies intercept ingoing and outgoing traffic to their associated agent.
 The service mesh controller also provides the proxies with identities for our agents as well as a key pair for them to interact securely via mutual TLS authentication (\texttt{mTLS}).
 \texttt{mTLS} provides authentication and encryption.
 Some examples of service meshes include Istio~\cite{istio} and linkerd 2.0~\cite{linkerd}.
 \item \textbf{Policy Engine}: A policy engine is a system answering policy-related queries according to a policy configuration.
 In the service mesh, every agent can communicate with all the other agents by default.
 We add a policy engine as a sidecar in each pod, so that the policy can be enforced.
 Each time a sidecar proxy intercepts a request from its associated agent, it checks with its policy sidecar if the request is authorized or not.
 If the request is authorized, the request is forwarded accordingly.
 If the request is not authorized, the proxy returns an error message to the service.
 Rules can be added to the policy to constrain the communications an agent can make.
 Additional rules can be used to enforce some other constraints based on context, like a time period during which a given communication is authorized to occur.
 A policy engine example is Open Policy Agent (OPA)~\cite{opa}.
\end{itemize}

\begin{figure}
 \centerline{\includegraphics[width=1.15\linewidth]{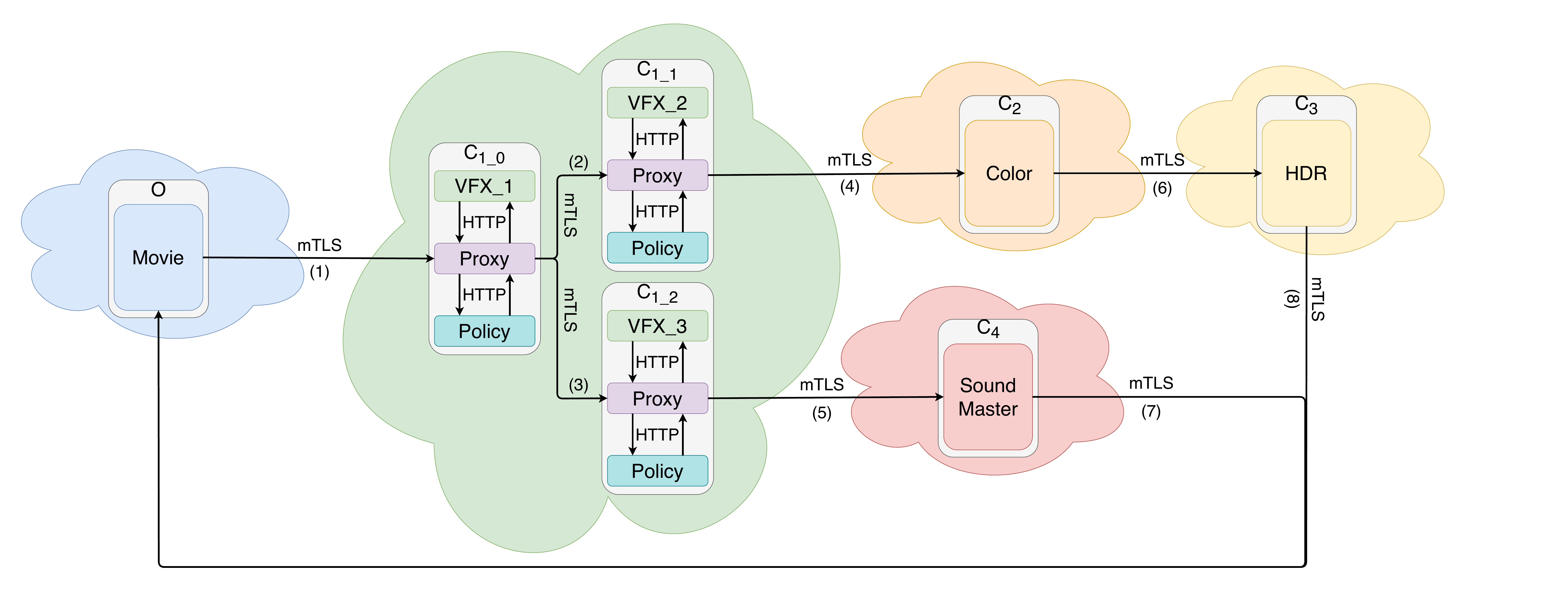}}
 \caption{Secure infrastructure. Each box represents an agent. It is a pod with the appropriate containers. The container of the color of the actor represents the service. Purple containers represent the proxies of the service mesh, and blue containers represent the policy sidecars. The arrows stipulate whether the communications are secure (\texttt{mTLS}) or not (HTTP).}
 \label{fig:wfs-3}
\end{figure}

Figure~\ref{fig:wfs-3} shows the workflow we defined in Fig.~\ref{fig:wfs}, with each actor having its own deployment space represented by the cloud surrounding the boxes which represent the agents of those actors (e.g. the $C_{1\_1}$ box represents an agent of contractor $C_{1}$).
The access policies of a service are pushed in the policy sidecar associated with the service.

\begin{figure}
 \centerline{\includegraphics[width=0.9\linewidth]{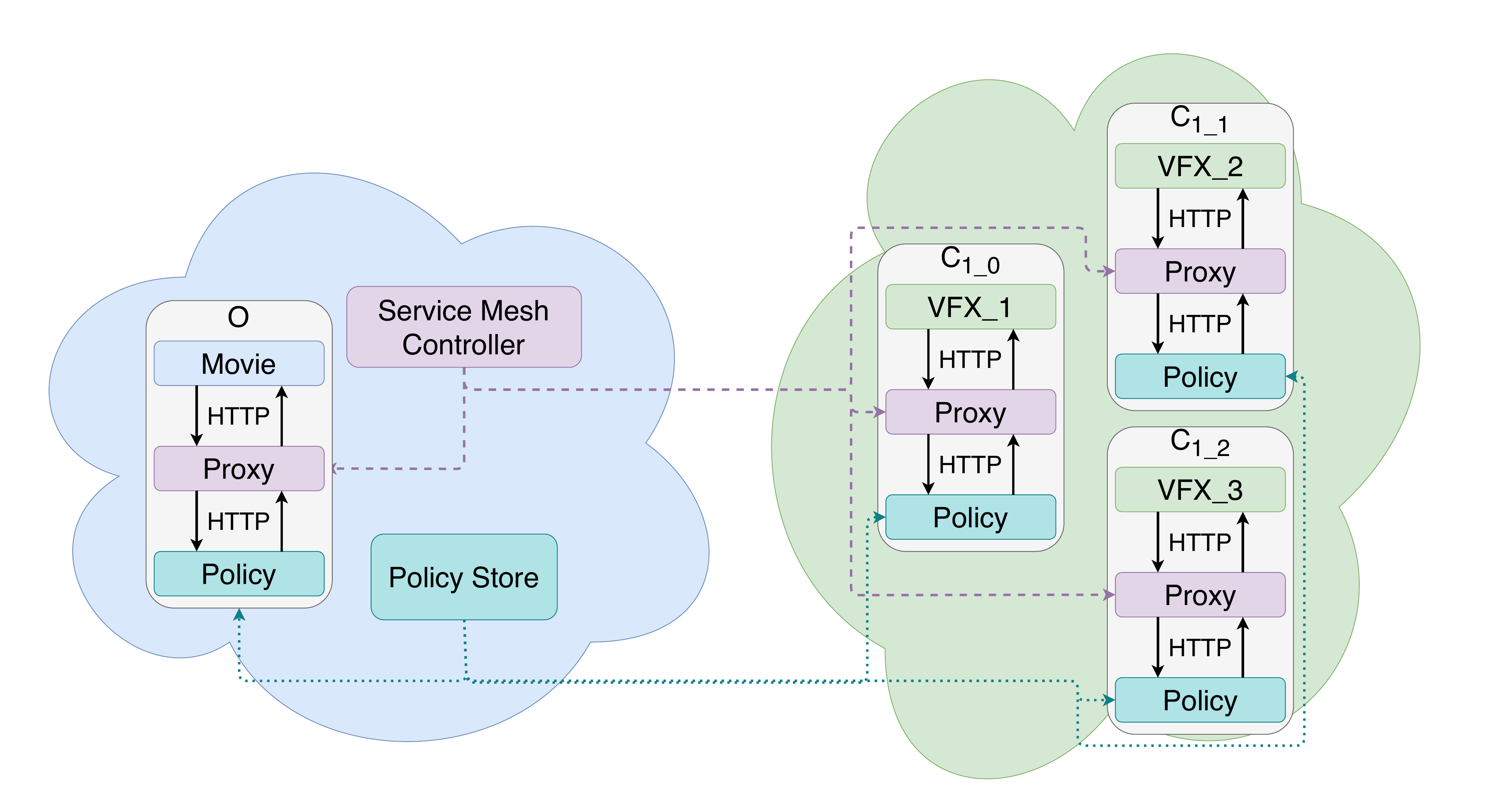}}
 \caption{Representative subset of the secure infrastructure control plane (contractors $C_2$ through $C_4$ are not represented). Proxies are configured by the service mesh controller, providing them with identities and key pairs, as well as routing information for them to initiate secure communications with other proxies in the mesh. Policy changes are enabled with periodical pull on the policy sidecars (whose input comes from a policy store).}
 \label{fig:wfs-3-ctrl}
\end{figure}

Figure~\ref{fig:wfs-3} also illustrates how we use the elements of the microservice architecture.
Each agent is a pod, containing the service (i.e., the environment the agent will be using), a proxy and a policy sidecar.
The proxy sidecar will intercept all traffic coming from and going to its respective service.
The proxy will then check thanks to the policy sidecar if the request is authorized or not.
If the request is authorized, it is forwarded accordingly, and the request is rejected otherwise.
Proxies are configured by the service mesh controller (Fig.~\ref{fig:wfs-3-ctrl}), providing them with identities and key pairs, as well as routing information for them to initiate secure communications with other proxies in the mesh.
Policy is pulled periodically by the policy sidecars from a policy store, which allows for policy changes.
Since the service mesh controller and the policy store are under the control of the owner, he is in control of the system.
Thus, the owner specifies the policy to be applied to enforce the desired workflow, preventing data from leaking outside.

The data processed by the pods is stored on mounted Persistent Volumes (PVs), which are encrypted with a key located in a key-value store of the orchestrator, providing us with \textbf{data security at rest}.
We generate a key to encrypt each PV required by needs of the workflow.
Since the keys are all stored in the same key-value store, this does not really mitigate risks against a technically skilled attacker gaining access to the key-value store, but it can help to protect some of the data in case the attacker only gains access to a subset of the keys through other means.
Since the keys are stored in the master components of the orchestrator, they are under the control of the owner.
To enforce the workflow and make sure the agents cannot bypass it via the PVs, each agent must have its own personal PV.

Pods can also communicate according to the specified workflow and policy via \texttt{mTLS}, providing us with \textbf{data security in transport} as indicated by the communications between the pods in Fig.~\ref{fig:wfs-3}.
Communications inside a pod are not encrypted, but the isolation layers protect the data against eavesdroppers.

Once a service has completed all the tasks he was assigned to do, the associated pod is destroyed to make sure data cannot be leaked from this service past this point in time.
To provide data security in transport, services in the service mesh are provided with an identity in the form of a certificate, which is associated with a key pair.
To make sure those are safe to use, and that no attacker gained access to the keys or tampered with them before they reach the appropriate service, we need to verify the key distribution process is secure.
In the case of the service mesh, this is done automatically for us.
Appendix~\ref{app:bootstrap} summarizes this procedure, and how it is secured.
Thanks to this infrastructure, communications can be constrained to follow a policy, giving us a streamlined way to prevent data leaks.
We show how a simple policy to prevent data leaks can be defined.

\subsection{Authorization is Challenging}
\label{sec:authz}

A solution to this problem needs to take into account that an ecosystem is composed of a diverse set of actors and agents~\cite{mehta2017}, operating on multiple types of resources, with multiple available operations.
This diverse environment means that the problem that we are trying to solve is very general: we need a way to determine whether or not an Identity $I$ can perform Operation $OP$ on Resource $R$, depending on attributes $A$, for all combinations of $I$, $OP$, $R$ and $A$.
We want a solution that considers the quadruplet ($I$, $OP$, $R$, $A$), else it would require to deploy multiple layered solutions, thus complexifying control and visibility over our authorization system.
In our case, it implies considering combinations of actors and agents ($I$) with all resources ($R$), all possible operations ($OP$), and applicable attributes ($A$).

For example, our workflow (Fig.~\ref{fig:wfs-3}), is composed of an owner and four contractors, with contractor $C_{1}$ having three agents.
In our example, our agents may interact with each other via the paths laid out in Fig.~\ref{fig:wfs-3}, only with the "write" operation on the specified path to access the resource, "/data".

To solve this problem of authorization, we need a way to specify rules, in order to enforce the policy of the data owner.
An authorization model tells us how to specify those rules.

For our needs, an ideal authorization model would be one that is:
\begin{itemize}
 \item \textbf{Fine-grained}: The level of detail that can be expressed with the authorization model is accurate enough to specify the needs of the owner in terms of policy.
 \item \textbf{Flexible}: The authorization model takes into account environment variables (e.g. time, mode of connection).
 \item \textbf{Centralized}: The policy specified in the authorization model is under the control of a single entity.
 In our case, this entity is the data owner.
 \item \textbf{Relational}: The authorization model is built on relations between the different components of the model (e.g. agents, roles, resources), making it easy to determine the permissions of an agent and/or the capabilities of a resource.
\end{itemize}

The four archetypes when it comes to authorization models are Mandatory Access
Control~\cite{sandhu1993lattice} (MAC), Discretionary Access Control~\cite{sandhu1994access} (DAC), Role-Based Access Control~\cite{david1992role} (RBAC), and Attribute-Based Access Control~\cite{hu2013guide} (ABAC).
Looking at those archetypes, a solution matching our requirements would be to use a hybrid model~\cite{coyne2013}.
In the role-centric variation of the model, attributes are added to constrain RBAC further, only reducing permissions granted to a subject.
This is a good compromise, as the RBAC model is preserved, and it gains in flexibility thanks to the added attributes.

In our example, a role-centric variation between RBAC and ABAC translates to first setting the permissions for our roles, in this case corresponding to our agents (e.g. $O, C1_{1}$), depending on the considered resources (e.g. the movie data) and operations (e.g. read, write, ...).
Those permissions can then be further constrained by adding requirements to the existing rules in the form of attributes (e.g. $C_{1\_2}$ can only push data to $C_{4}$ before a specified deadline).

\subsection{Formalization of the Example Workflow for Access Control}

As a simple example, we will formalize the workflow of Fig.~\ref{fig:wfs-3} with only a write operation, one resource and one attribute.
To formalize the workflow of Fig.~\ref{fig:wfs-3} for access control purposes, we first need to define the identities of the workflow: $I = \{O, C_{1\_0}, C_{1\_1},\allowbreak C_{1\_2}, C_{2}, C_{3}, C_{4}\}$.
Since data will only be pushed along the workflow, we only need a $write$ operation: $OP = \{W\}$.
One resource is to be considered, the movie modified by the contractors: $R = \{R_{movie}\}$.
Lastly, we need some attributes to constrain access control permissions, like the date of access: $A = \{A_{date}\}$.

We thus have the following:
\begin{gather*}
I = \{O, C_{1\_0}, C_{1\_1}, C_{1\_2}, C_{2}, C_{3}, C_{4}\} \\
OP = \{W\} \\
R = \{/data\} \\
A = \{A_{date}\}
\end{gather*}

To construct the access control policy, we can transform the graph of a workflow into the policy by first numbering the edges of our workflow graph (Fig.~\ref{fig:wfs-3}), and then constructing the associated policy by adding a permission for each numbered edge in our graph.
Listing~\ref{lst:ac-ver1} shows this policy.
The edge number one of Fig.~\ref{fig:wfs-3}, corresponds to the first line of $role\_perms$ (line 4), and so on.

This policy can be further defined by adding time constraints, for example if $C_{1\_2}$ can only push data to $C_{4}$ before a specified deadline.
Listing~\ref{lst:ac-ver2} shows this policy.

Looking back at our attacker model, we can see how the elements of the infrastructure prevent them from attaining their goal.
Since services can only be reached by going through the proxy (and its associated policy sidecar), the third-party attacker cannot perform any data breach.
The co-located attacker faces the same issue, and system exploits may be prevented via the isolation provided by pods and containers.
The malicious contractor agent cannot leak data outside of the workflow, since the implemented policy prevents him from sending data anywhere else.
Obviously, protection is only provided if the trust model is respected, and the policy is defined and implemented correctly.
The owner has all the tools in hand to do so, since the service mesh controller and the policy store are under its control.

\section{Proof of Concept}
\label{sec:poc}

We realized a proof of concept, by implementing the infrastructure described in Sec.~\ref{sec:descr}.
We reproduce the workflow of Fig.~\ref{fig:wfs-3}, with services of the workflow receiving and sending arbitrary data to represent the data of the owner.
We use Docker~\cite{docker} for our containers, Kubernetes~\cite{kubernetes} for our orchestration layer, Istio~\cite{istio} as our service mesh, using Envoy~\cite{envoy} for the proxy sidecars and Open Policy Agent~\cite{opa} for the policy sidecars.
We also use Kubernetes to provide the services with encrypted volumes.
This infrastructure was deployed on Google Cloud Platform (GCP), using one cluster for each actor of the workflow, for a total of five clusters.
Each cluster runs one \textbf{n1-standard-2} node (2 vCPUs, 7.5 GB of memory), on version \textbf{1.14.10-gke.36}, except the cluster of the owner which runs two of them, since running the control plane requires additional resources.
The clusters for the owner, color and VFX are located in us-central1-f whereas the clusters for HDR and sound are located in us-west2-b.

The workflow we want to enforce is shown in Table~\ref{tbl:ac}, where each row represents the \textbf{source} of a request, and each column a \textbf{destination}.
The full policy is not represented by this table, additional attributes further constrain those permissions (see Listing~\ref{lst:ac-poc} for the full policy).
The agents can also send GET requests, but they are all denied by the policy.

\begin{table}[ht]
\centering
\caption{Proof of Concept policy}
\resizebox{0.6\columnwidth}{!}{
\begin{tabular}{lccccccc}
\toprule
{}	     & \multicolumn{7}{c}{Destination}\\\cmidrule{2-8}
Source   & owner & $VFX_1$   & $VFX_2$ & $VFX_3$ & Color & Sound & HDR  \\\midrule
owner    & -     & POST      &         &         &       &       &      \\
$VFX_1$  &       & -         & POST    & POST    &       &       &      \\
$VFX_2$  &       &           & -       &         & POST  &       &      \\
$VFX_3$  &       &           &         & -       &       & POST  &      \\
Color    &       &           &         &         & -     &       & POST \\
Sound    & POST  &           &         &         &       & -     &      \\
HDR      & POST  &           &         &         &       &       & -    \\
\bottomrule
\end{tabular}}
\label{tbl:ac}
\end{table}

The complete data, code as well as guidance to realize this Proof of Concept are publicly available\footnote{See github.com/xxx. We will publish the repository upon acceptation of the paper.}.

We also developed a test framework to check that:
\begin{itemize}
 \item Traffic is either encrypted or protected inside a pod by the isolation provided by the pods;
 \item The policy, allowing or denying communications between services, is correctly enforced.
\end{itemize}

To do so, we capture traffic on every network interface in the service mesh and perform each possible communication.
In the general case, considering we have $N$ services and $M$ types of request, we obtain the number of possible communications with the formula: $N(N-1)M$.
Since we capture on each interface, and services have a loopback as well as an external interface, we obtain the total number of required captures: $N(N-1)M(2N) = 2(N^3 - N^2)M$.
The number of required captures thus grows cubicly with the number of services and linearly with the number of requests.

\begin{figure}[htbp]
 \centerline{\includegraphics[width=0.8\linewidth]{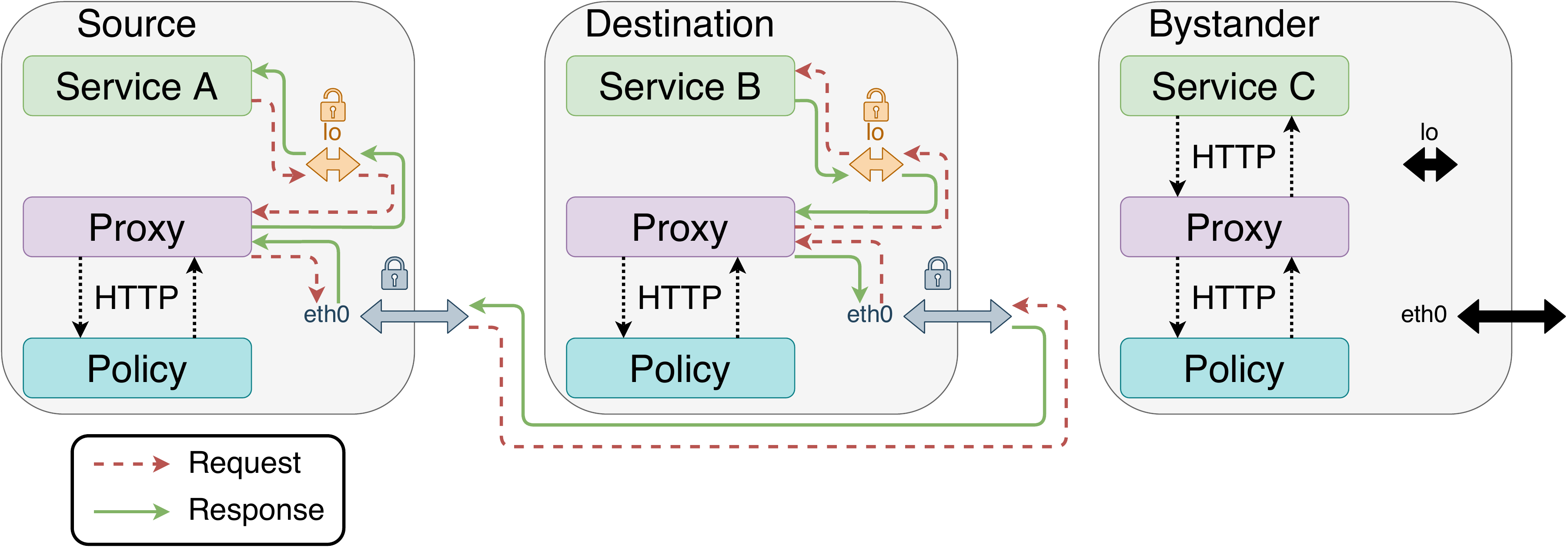}}
 \caption{Detailed view of pods and the communication flow. Traffic is unencrypted on the loopbacks, but encrypted on the external interfaces.}
 \label{fig:testing}
\end{figure}

Considering our previous example in Sec.~\ref{sec:poc}, we have seven services, two possible requests (GET and POST), which gives us a total of 1176 captures.
Captures are obtained from a \texttt{tcpdump} container added to the service pods as a sidecar.
Since containers in the same pod share the same network namespace, capturing traffic from the \texttt{tcpdump} container on either the loopback or the external interface allows us to see traffic from the other containers in the pod.
Figure~\ref{fig:testing} shows the path a communication takes inside the service mesh, as well as whether traffic is encrypted or not.
The request is initiated by service A, and intercepted by its associated proxy via the loopback interface.
The proxy will then check thanks to the policy sidecar if the request is authorized or not.
If the request is authorized, it is forwarded accordingly.
The request is rejected otherwise.
In the case where the request is authorized, it is forwarded to the proxy of service B by using \texttt{mTLS}.
There, the proxy forwards the request to service B, which replies by going through the same steps as earlier.
Traffic going to/coming from the loopback should be unencrypted, whereas traffic going to/coming from the external interface should be encrypted.
Our captures show that this is indeed the case.
Traffic does not need to be encrypted on the loopbacks, as all the elements (i.e, the service and its sidecars) that have access to this loopback are in the same trust zone.
The layers of isolation provided by the pods protect the loopback traffic from being seen by unauthorized entities.

Obviously, pods in the service mesh have one of three roles during a communication.
Either they are the source of the communication, the destination of the communication, or simply a bystander that is not involved in the communication.
This is important, because the checks we need to perform depend on where traffic was captured:
\begin{itemize}
 \item \textbf{Source/Destination loopback}: We need to verify that a communication between the source and the destination is occurring (i.e., correct IP addresses and ports).
 We need to verify that the request in the capture corresponds to the request we are testing for (GET or POST).
 The response needs to be in accordance with the policy: in this case, '403 Forbidden' if the policy was 'deny' and '200 OK' (GET) or '201 OK' (POST) if the policy was 'allow'.
 \item \textbf{Source/Destination external interface}: We need to verify that a communication between the source and the destination is occurring (correct IP addresses and ports).
 We need to verify that the traffic is encrypted by \texttt{mTLS}, and not passed in clear text.
 \item \textbf{Bystander loopback and external interface}: We need to verify that no communication between the source and the destination is occurring, whether encrypted or unencrypted.
\end{itemize}

We built a tool that automatically extracts the authorization policies from the OPA policy configuration, generates and then tests an access control matrix.
For each possible communication in the service mesh, our tool loads all the captures relevant to this communication, identifies them to see what we should verify in each capture, and then proceeds to check if captures are in accordance with the criteria above.
It is then easy to evaluate whether the system is compliant with the overall policy.
The complete code for the test framework is publicly available\footnote{See github.com/xxx. We will publish the repository upon acceptation of the paper.}.

\section{The Overhead of Security}
\label{sec:measures}

In this section, we analyze the performance overhead added by the policy sidecar enforcing security.
We measure the pod startup time, and the request duration (between each couple of connected pods).

\subsection{Startup time}

We first evaluate the impact of having an additional container for OPA on the startup time of pods.
An independent-samples t-test was conducted to compare startup times in a deployment of our PoC with or without OPA. 
Our aim is to determine whether these is a significant statistical difference in startup time when OPA is instantiated. 
For that purpose, we test two hypotheses. 
($H_0$) There is no statistically significant difference in the startup times of the two deployments, and ($H_1$), a statistically significant difference exists between the startup times of the two deployments.

We gathered 130 observations per pod per deployment ($N = 1820$ total). We measure the startup time of pods by scraping the transition timestamps between \texttt{PodScheduled} and \texttt{Ready} states of the pods. 

\begin{figure}
 \centerline{\includegraphics[width=0.75\linewidth]{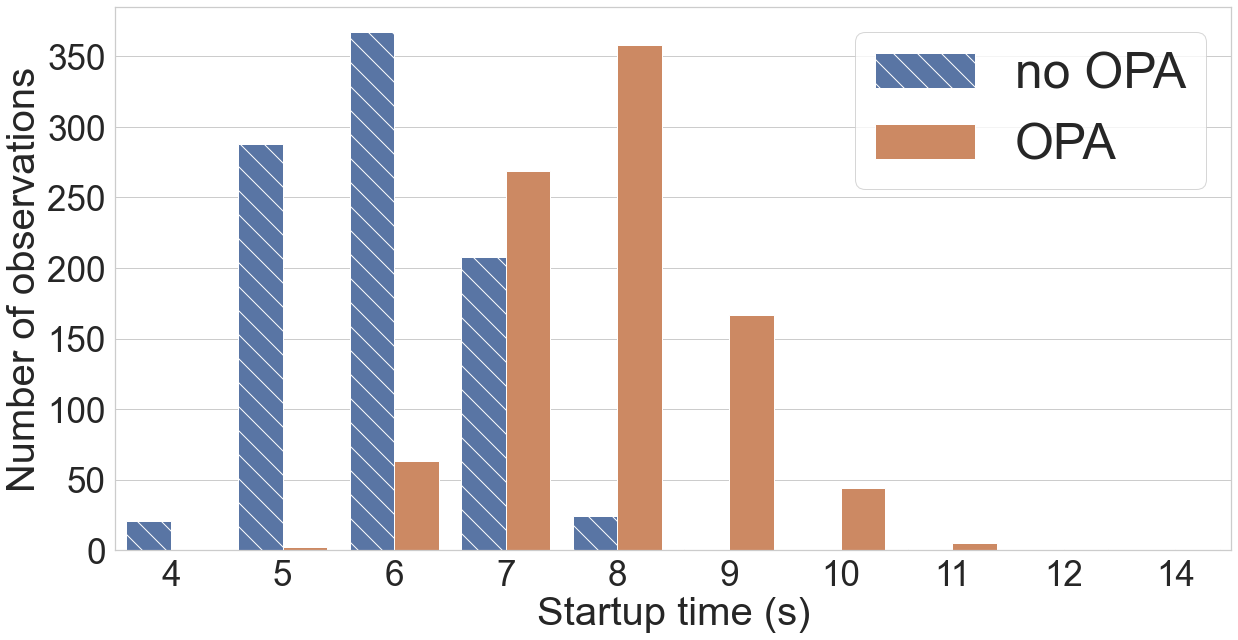}}
 \caption{Distribution of startup time in the deployment without OPA (stripes) compared to the deployment with OPA (full)}
 \label{fig:startup-time}
\end{figure}

Figure \ref{fig:startup-time} allows to compare the distribution of startup times in the deployments with and without OPA to measure the cost on the initial deployment.

The 910 observations in the group with the OPA sidecar ($M = 7.87, SD = 1.03$), compared to the 910 observations in the group without the OPA sidecar ($M = 5.93, SD = 0.88$), exhibit significantly higher startup times, $t(1818) = 43.19, p < 0.001$.
The \textit{effect size} for this analysis, $d = 1.985$, was found to exceed Cohen's convention for a large effect ($d = 0.80$).
Running a \textit{post hoc power analysis} also reveals a high statistical power, $1 - \beta > 0.999$.

These results show that pods with OPA have a substantial increase in startup time of almost two seconds on average.
This corresponds to an increase of 32.72\% of the startup time.
We ran the same analysis on a per pod basis and found the results to be consistent with those results.

\subsection{Request time}

This second evaluation is about measuring the influence of the policy size on communications, to test whether the policy is scalable for more complex workflows.
We analyze separately requests happening within the same region (intra-region e.g. us-central1-f to us-central1-f) from inter-region requests (e.g. us-central1-f to us-west2-b) as communication times are significantly different in the two cases.

For each request duration between pairs of pods, we take into account whether the authorized communication happened inside a region or between regions (inter-region), since the durations of requests are different for those types of communication, and thus the impact of policy size is not the same.

A \textit{one-way between subjects ANOVA} was conducted for each type of communication (intra-region and inter-region) to compare the effect of policy size on request duration in five levels of policy size: \texttt{no opa}, \texttt{all allow}, \texttt{minimal}, \texttt{+100} and \texttt{+1000}.

The \texttt{no opa} policy deployment corresponds to having no OPA container at all.
The \texttt{all allow} policy deployment corresponds to having no rules and allowing all communications by default.
The \texttt{minimal} policy deployment corresponds to having the default minimal number of rules to enforce the workflow of the PoC.
The \texttt{+100} and \texttt{+1000} correspond to the minimal policy being inflated respectively with 100 additional rules ($+147\%$) and 1000 additional rules ($+1470\%$), with additional rules being obligatorily evaluated by OPA.

For each ANOVA, we gathered 40 observations per authorized communication per level of policy ($N = 1600$ in total).
We measure request duration by recording the elapsed time of requests between pairs of pods via cURL.

\begin{figure}
 \centerline{\includegraphics[width=0.75\linewidth]{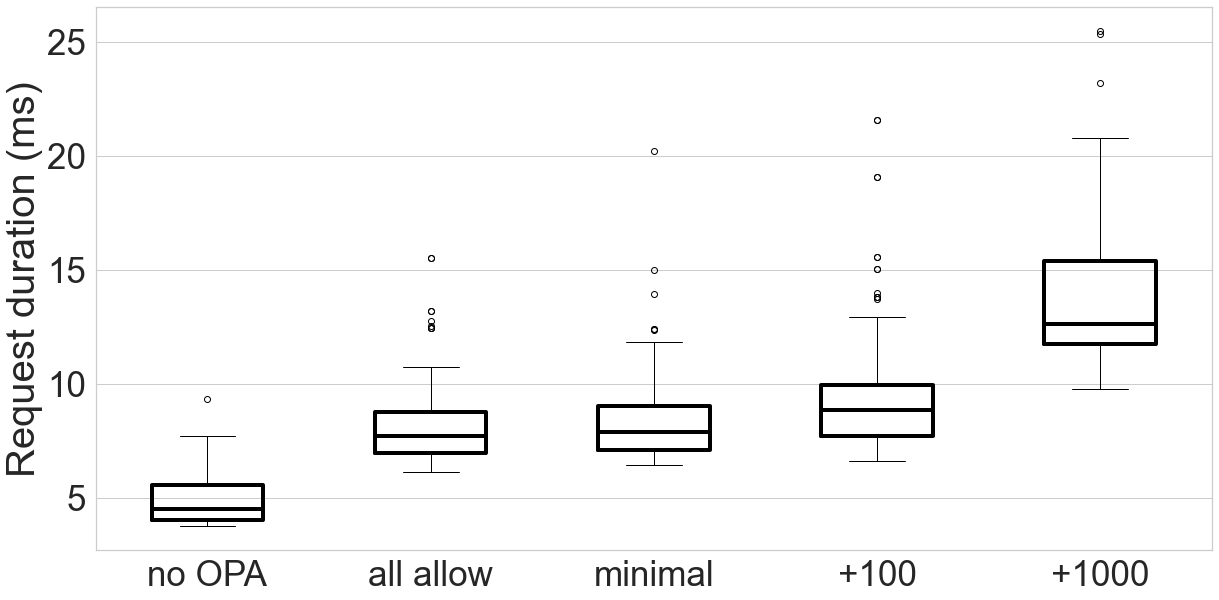}}
 \caption{Spread of request duration in intra-region communications by policy size}
 \label{fig:request-duration}
\end{figure}

Figure \ref{fig:request-duration} shows the distribution of request duration for each policy size. 
For intra-region communications, there is a significant difference in request duration among the five scenarios of policy deployments, $F(4, 795) = 364.05, p < 0.001, {\eta^2}_p = 0.65$.
Post hoc comparisons using Tukey's HSD test indicated significant differences between the mean scores of all policy deployments, e.g. \texttt{no opa} ($M = 0.0049 s, SD = 0.0010$), \texttt{+1000} ($M = 0.0136 s, SD = 0.0028$), except the \texttt{all allow} and \texttt{minimal} deployments.

For inter-region communications there also exists a significant difference (in request duration) among the five scenarios of policy deployments, albeit with a lesser effect: $F(4, 795) = 15.23, p < 0.001, {\eta^2}_p = 0.07$.
Post hoc comparisons using Tukey's HSD test indicated that the mean scores for policies close in size were not significantly different (e.g. \texttt{no opa} and \texttt{all allow}), whereas policies distant from each other in terms of size were found significantly different (e.g. \texttt{+100} and \texttt{+1000})\footnote{See github.com/xxx for full data, code and statistical analysis in the form of jupyter notebooks. We will publish the repository upon acceptation of the paper.}.

As expected, the impact of policy size is weaker in the inter-region communications as those communications are naturally longer and more subject to variations than intra-region communications.
Policy size accounts for 65\% of the variance in intra-region communications whereas it accounts only for 7\% of the variance in inter-region communications.

Taken together, these results suggest that policy size does have a considerable effect on request duration.
While our results suggest that a higher policy size increases request duration, it should be noted that the size must be quite high in order to observe an effect, especially regarding inter-region communications.
With such communications, incremental increases in policy size do not appear to have a significant effect on request duration.

\section{Discussion}
\label{sec:discuss}

Our infrastructure enables the secure exchange of data and its safety at rest while avoiding any leak.
More specifically, in accordance with the principles of zero-trust, we achieve a secure system that enables the exchange of data between non-trusted agents while guaranteeing this data is secure at rest, in transport and cannot be leaked by any agent in both cases, in the context of a workflow.
With our model and PoC, we provide \textbf{data security at rest} by using encrypted volumes as well as multiple layers of isolation in the form of pods and containers.
We also provide \textbf{data security in transport} by using mTLS for the communications that are exposed, which provides us with encryption, authentication and integrity of the data being exchanged.
The workflow is defined by the owner and enforced using policy sidecars, which controls the agents participating in the workflow.

Looking back at our attacker model, one can see how the elements of the infrastructure prevent adversaries from attaining their goal.
The malicious agent cannot leak data outside of the workflow, since he is isolated in a pod, and all requests emanating from the agent are intercepted and submitted to the implemented policy, which prevents him from leaking data.
Communications the agent makes are encrypted, and can only be performed with other authenticated agents, preventing eavesdropping and impersonation.
Since services are isolated and can only be reached by going through the proxy (and its associated policy sidecar), an external attacker cannot perform any data breach.
The co-located attacker faces the same issue, and system exploits may be prevented via the isolation provided by pods and containers.
Obviously, protection is only provided if the trust model is respected, and the policy is defined and implemented correctly.
The owner has all the tools in hand to do so, since the service mesh controller and the policy store are under its control.

Using our solution comes with other practical advantages.
Our infrastructure is easy to use as each actor can opt for the environments of its choice in order to deploy the workflow.
The standardized environment grants reduced complexity for security management, compared to monolithic solutions.
Changes to the workflow and the policy can be applied at the centralized policy store, using a high-level policy language.
The business intelligence of the actors is separated from the infrastructure; since our architecture model is modular, one can choose which solution to use for each part of the infrastructure.
A service mesh can be swapped for another, same thing for the orchestrator and the container technology, meaning there is no vendor lock-in, the solution is flexible.

\section{Related Work}
\label{sec:rel}

Existing works provide guidance on overall security requirements and strategies for microservices~\cite{chandramouli2019security}, as well as guidance on more specific microservices components like service mesh~\cite{el2019guiding,chandramouli2020building} or containers~\cite{souppaya2017application,chandramouli2017security}.
Chandramouli~\cite{chandramouli2019security} provides guidance on security strategies for implementing core features of microservices, as well as countermeasures for microservices-specific threats.
El Malki and Zdun~\cite{el2019guiding} provide guidance on service mesh based microservice architecture decisions by studying existing practices.
Chandramouli and Butcher~\cite{chandramouli2020building} give recommendations for the deployment of service mesh components in the proxy sidecar version of the microservice architecture.
Souppaya et al.~\cite{souppaya2017application} provide a listing of major security risks associated with the use of containers, as well as recommendations to counter those risks.
Chandramouli~\cite{chandramouli2017security} then analyzed different security solutions for containers, to see if they were in accordance with recommendations of~\cite{souppaya2017application}.

Weever et al.~\cite{de2020zero} investigate operational control requirements for zero-trust network security, and then implement zero-trust security in a microservice environment to protect and regulate traffic between microservices.
Hussain et al.~\cite{hussain2019intelligent} propose and implement a security framework for the creation of a secure API service mesh using Istio and Kubernetes.
They then use a machine learning based model to automatically associate new APIs to already existing categories of service mesh.
Zaheer et al.~\cite{zaheer2019eztrust} propose eZTrust, a policy-driven perimeterization access control system for containerized microservices environments.
They leverage eBPF to apply per-packet tagging depending on the security context, and then use those tags to enforce policy.

On the side of formal analysis of data leaks in workflows, Accorsi and Wonnemann~\cite{accorsi2011strong} proposed a framework for the automated detection of leaks based on static flow analysis by transforming workflows into Petri nets.
Khan et al.~\cite{khan2019data} propose a synthesis on data breach risks and resolutions, identifying and classifying data breaches, locus and impact, and then giving guidance on prevention, containment and recovery.

Regarding policy, Ranathunga et al.~\cite{hamza2018clear,hamza2019verifying} propose a tool to help manufacturers in the development and verification of formal profiles for IoT devices, modeling network access control policies using metagraphs, and then checking their compatibility with a given organizational policy as well as best practices.

\section{Conclusion}
\label{sec:conclusion}

In this work, we described and implemented a secure architecture that prevents data leaks and protects business intelligence.
More specifically, in accordance with the principles of zero-trust, we achieve a secure system that enables the exchange of data between non-trusted agents while guaranteeing this data is secure at rest, in transport and cannot be leaked by any agent in both cases, in the context of a workflow.
We realized a proof of concept of our architecture, monitored key parts of the workflow to show how the data is secured.
In addition, our experiments show that our approach scales to increasing workflow complexity.
We reviewed how our infrastructure was resilient to the set of attacks considered in our security model and discussed its benefits.

In the future, we plan to study how changes in the workflow impact the security of the system.
Some work could also be done on the removal of trust requirements, by adding Trusted Execution Environments to our infrastructure, or also using Smart Contracts.
This would provide us with a fully secure environment, so that even an actor with administration rights on the machine cannot peer at the data.
Even though the data in our infrastructure is encrypted at rest, this would also give us another guarantee that the data of the owner and the business intelligence of the contractors are secure for any processing task.
%
%
%
\bibliographystyle{splncs04}
\bibliography{secure-workflow}

\appendix

\section{Full Removal of Trust is not an Option?}
\label{sec:trust}

In this section, we explore the possibility of changing our trust model. We remove the favorable assumption we made that the owner and the contractors trust each other. Indeed, without this assumption, there is no easy way for the owner to verify that the data he is sending to a contractor is actually (and only) going to the secure environment we will present in Section~\ref{sec:descr}, therefore removing any guarantees one might have concerning data leaks.
A contractor might replicate and redirect incoming workflow traffic to an insecure location and decrypt it using the keys the owner provided them with to achieve secure communication.
In this case, the workflow data is stored insecurely and might be leaked to third parties.
We also have no guarantees that the contractors are actually applying any policy that the owner wants to enforce.

\begin{figure}[htbp]
 \centerline{\includegraphics[width=0.6\linewidth]{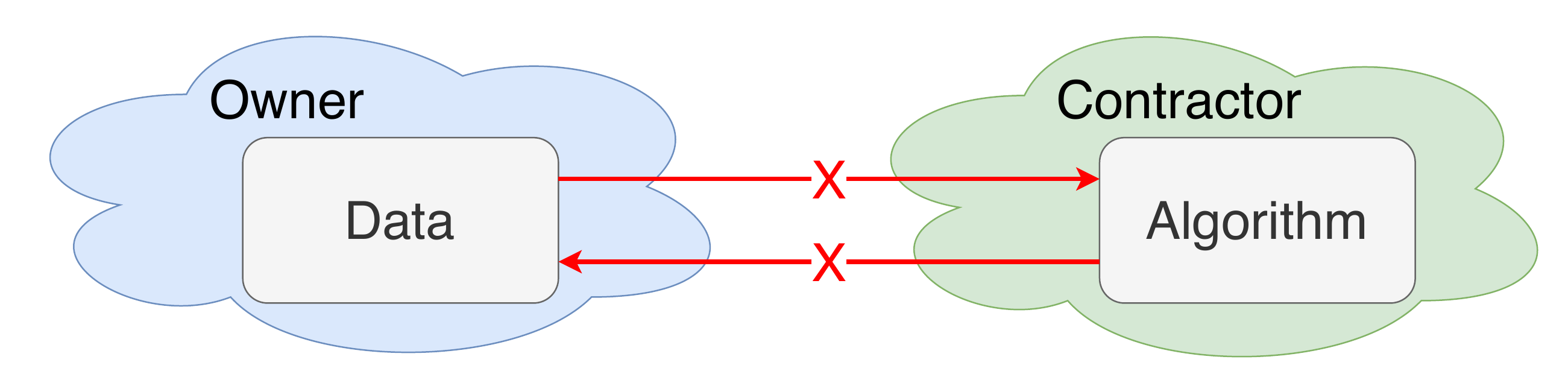}}
 \caption{Problem arising when we remove trust between actors. Nor the owner can send its data to the contractor, as it is no longer trusted, neither the contractor can send its business intelligence / algorithm to the owner, for the same reasons.}
 \label{fig:dist-syst-problem}
\end{figure}

In such a situation, to realize the workflow, or in other words, to perform the tasks on the data of the owner, we cannot send the data of the owner to the contractors, as they are no longer trusted, but we also cannot send the business intelligence of the contractors to the owner for the same reasons (Fig.~\ref{fig:dist-syst-problem}).

Moreover, the tasks of the contractors might require some interaction with the agents of the contractors in the form of a comprehensive feedback loop, in particular with employees working on the special effects of a movie as they need to understand how their visual effects impact the movie.
This is a fundamental constraint since it means the processed data needs to be sent back to the contractors at some point for clear viewing.

In this case, a non-solution would be to send the data and the processing tasks of the contractor to some third-party application that will compute the result, but this is just a displacement of the issue, since both the owner and the contractors have to trust this third-party.
In the event this third-party is malicious, it can leak the data and the business intelligence of the contractor.

\begin{figure}[htbp]
 \centerline{\includegraphics[width=0.6\linewidth]{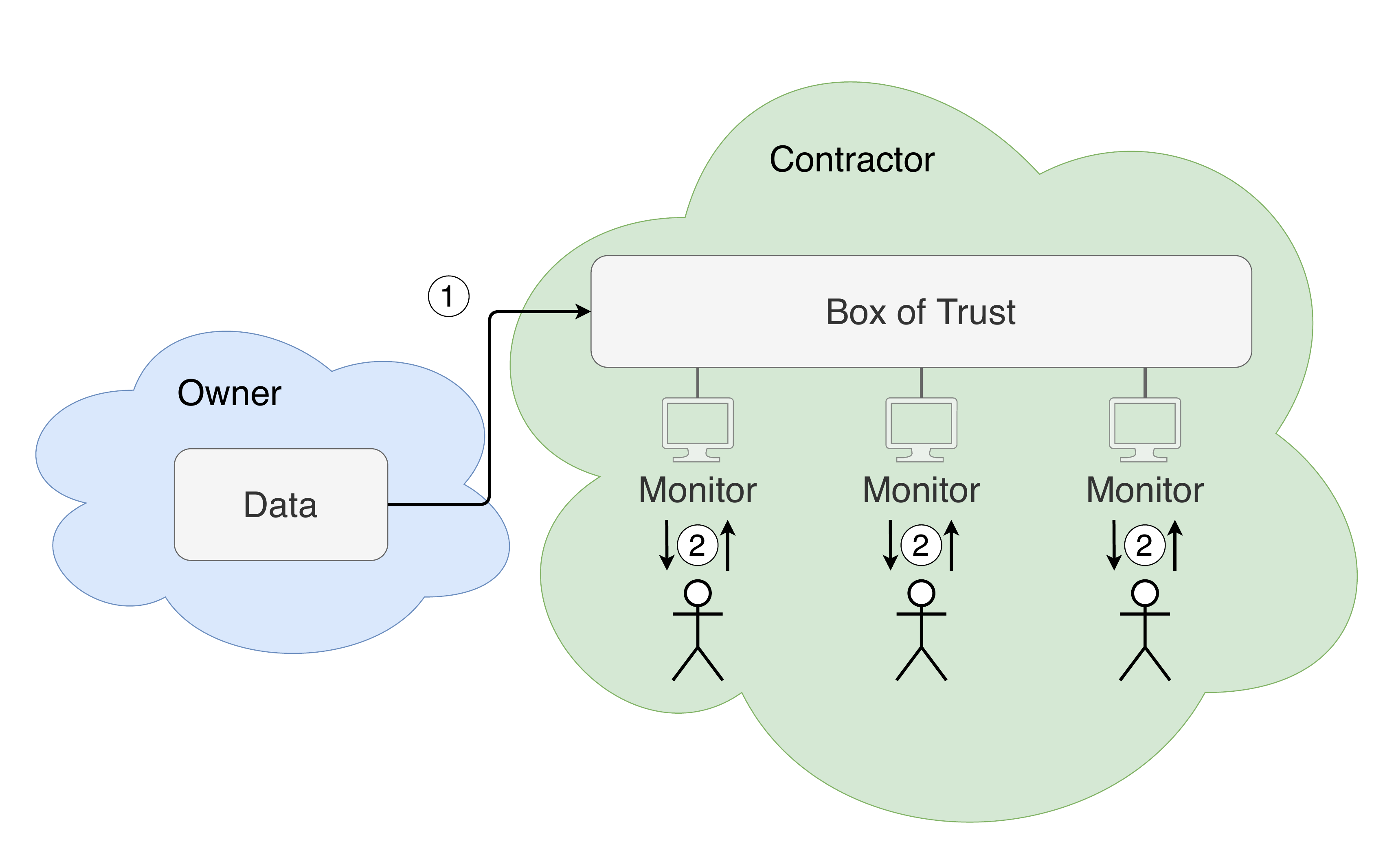}}
 \caption{Scenario 1 - The owner sends its data to BoT (Step 1), which are used by the contractor agents to fulfill their task(s) (Step 2).}
 \label{fig:dist-syst-box-ver1}
\end{figure}

Instead of trusting each other, actors can trust a third-party processor manufacturer that controls, at the hardware level, the access to resources (with a trusted computing architecture), or rely on open and audited hardware such as in the CrypTech project.
In other words, faced with this inherent limitation of data leaving a trust zone, the actors need to put their trust in a physical equipment, a trusted hardware device using TEEs.
The owner can then load its data in those physical equipments that we call BoT for Box of Trust, which are deployed by the contractor.
The contractor agents can then interact with the data to realize their tasks via input devices and monitors that are a part of these boxes (Fig.~\ref{fig:dist-syst-box-ver1}).
These boxes are provided with a set of private keys directly plugged in the physical secure environment to manipulate encrypted data at rest and in transport, and destroy the data and keys along with itself if one tries to tamper with it physically.
To load its business intelligence on the BoT, the contractor needs to have access to it.
As it is necessary -- but difficult -- to verify the data does not leave the trusted environment, one should verify the resources loaded by the contractor cannot leak the data outside its trusted environment. Such a verification is not trivial
and is still an active research topic.

In this first scenario, the BoT handles the computational load required to fulfill the tasks of the contractor, and is verified by the manufacturer, which gives a guarantee that the box has not been tampered with, and is as advertised.
In this case, the contractor never has access to the data except via feedback on the screens; that is the contractor does not fully own and pilot anymore its own reconfigurable hardware devices, screens included. Unless the contractor externally records the screens with a perfect camera, the data of the owner cannot leak in its original numerical format.
Finally, the owner never receives any information on the tools and methods the contractor used to fulfill its task, so the business intelligence of the contractor cannot leak, it remains in the box.

However, the contractors do not have anymore the opportunity to negotiate the physical hardware to use, including their screens.
They cannot decide on their own what devices they aim to use and they do not anymore fully possess their own infrastructure.
Moreover, this first scenario may not scale with large amounts of data and processing.
A BoT capable of handling the required computational load with good performances can be quite expensive, if not unrealistic.

\begin{figure}[htbp]
 \centerline{\includegraphics[width=0.6\linewidth]{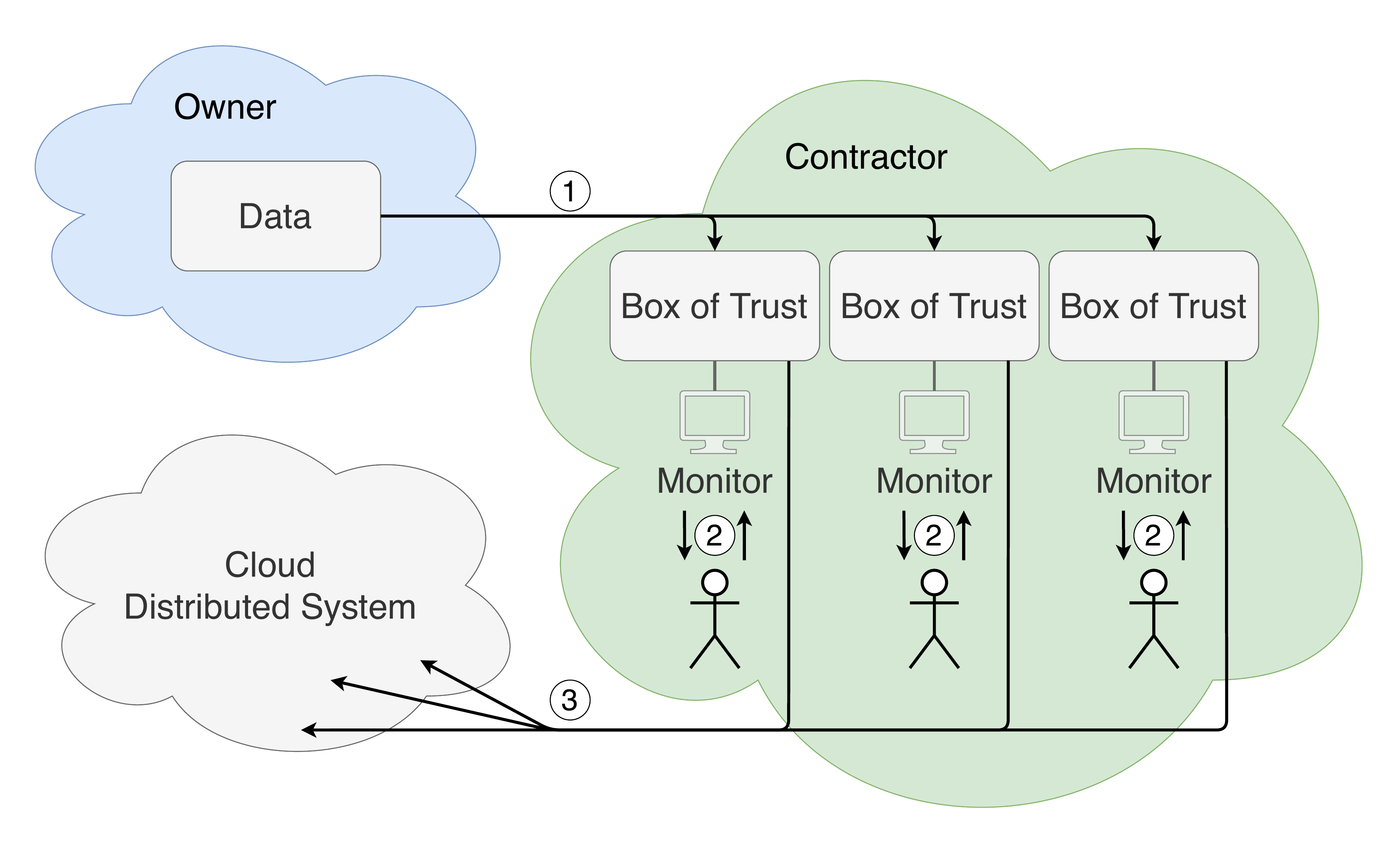}}
 \caption{Scenario 2 - The owner sends its data to BoT (Step 1), which are used by the contractor agents to fulfill their task(s) (Step 2). Operations and partial data are sent to a distributed system in the cloud to perform the operations the agents specify (Step 3).}
 \label{fig:dist-syst-box-ver2}
\end{figure}

To partially mitigate those issues, another scenario, shown in Fig.~\ref{fig:dist-syst-box-ver2}, makes use of lightweight BoTs for each agent and a cloud distributed system tasked with the execution specified by the agents via the BoT.
The owner once again loads its data in the BoT, and the contractor agents can interact with the data to realize their tasks via the monitors.
When an agent performs an operation, the BoT sends the part of the data the operation is realized on, along with the operation to realize to the distributed system, and the cloud executes the operation, returning the result to the box for display.

Since parts of the data and the operations of the contractor are sent in the cloud, we need to make sure the data as well as the business intelligence of the contractor are secure.
There are many ways to go about this:
\begin{itemize}
 \item \textbf{Load balancing}: Requests are load balanced across all machines.
 We need to trust the load balancing node is honest.
 Since the partial data and operation might end up on a malicious node, we need to split our data as well as the operations of the contractor.
 Thus, we need to determine at which granularity we want to partition the data and the algorithm of the contractor.
 The granularity can be determined by weighing the risk of partial data leak against performance: smaller chunks have better leak protection but worse performance, and vice versa.
 Ultimately, the granularity is determined by the owner and the contractor, by compromising between data leak protection and performance.
 To be sure that the malicious nodes of the distributed system cannot collude to discover the full data or the full operation the contractor wants to perform, we need to make sure the nodes receiving the partial data and operations cannot reach a consensus.
 Thus, as with the famous Byzantine Generals Problem, we need to trust that at least two thirds of the nodes in the cloud are honest.
 We could also use multiple clouds to reduce the risk of collusion in the cloud.
 \item \textbf{Fully Homomorphic Encryption}: Fully Homomorphic Encryption (FHE), which is able to perform arbitrary computations on encrypted data, could be used to process the data of the owner securely in the cloud.
 With FHE, there is no need to split the data again, as it is encrypted, but the operations realized by the contractor must be split similarly as in the load balancing scenario.
 In this case, no leaks are possible, but the process is very slow.
 Additionally, since data is encrypted but the operations are not, there is still a possibility one could retrieve the business intelligence of the contractor.
\end{itemize}

Since the cloud must for each operation send the results back to the contractor so that its agents can benefit from the full feedback loop, this scenario also exhibits performance limitations and does not come cheap (as with the first option), but is more flexible.
In this scenario, only the lightweight BoTs are required, as the BoTs act as a simple multiplexing point with a visual feedback loop.

Those scenarios aren't without drawbacks.
First, we are concerned with the decreased performance induced by TEEs.
Using TEEs incur decreased performance due to transitions, synchronization and paging~\cite{weichbrodt2018sgx}, especially if there is a lot of data to process.
In the case of Intel SGX, an enclave only has 128MiB of memory available (93MiB usable~\cite{weichbrodt2018sgx}), so each time the page needs to be swapped, it needs to be encrypted, and then paged out to memory.
The next enclave can then be paged in, and decrypted.
All those cryptographic operations lead to a performance hit.
Moreover, in both scenarios, contractors must use new dedicated equipment that comes at a cost, and can only use this equipment for the workflow.
This means control over the equipment is taken from the contractor, which leads to decreased deployment flexibility.
Having a third-party verify the resources loaded by the contractor to the BoT, and assert that those resources cannot leak the data, is also somewhat unrealistic.
Finally, akin to our infrastructure, the data is still vulnerable to physical attacks (e.g. an employee recording the screen).
Overall, making up for the removal of trust between actors to ensure the workflow can still operate securely and without leaks is costly, in terms of both performance and flexibility.
We argue that this tradeoff is not efficient and even unrealistic.
In the following, we then assume that organizations of the workflow trust each other and are considered honest but curious, whereas agents are assumed to be malicious (Sec.~\ref{sec:problem}).

\section{Identity and Authentication Bootstrap}
\label{app:bootstrap}

Since we use a service mesh, services are provided with an identity (certificate) which is associated with a key pair.
This identity and this key pair are then used to provide data security at rest, and data security in transport (authentication and encryption) via \texttt{mTLS}.
Trust in this identity, and the associated authentication mechanism is built on multiple layers, from the service mesh level down to the platform our infrastructure is deployed on.
In other words, secure communications at the service mesh level are securely bootstrapped by relying on secure communications at the orchestrator level, which are themselves securely bootstrapped by relying on secure communications at the platform level.

\begin{figure}[htbp]
 \centerline{\includegraphics[width=0.6\linewidth]{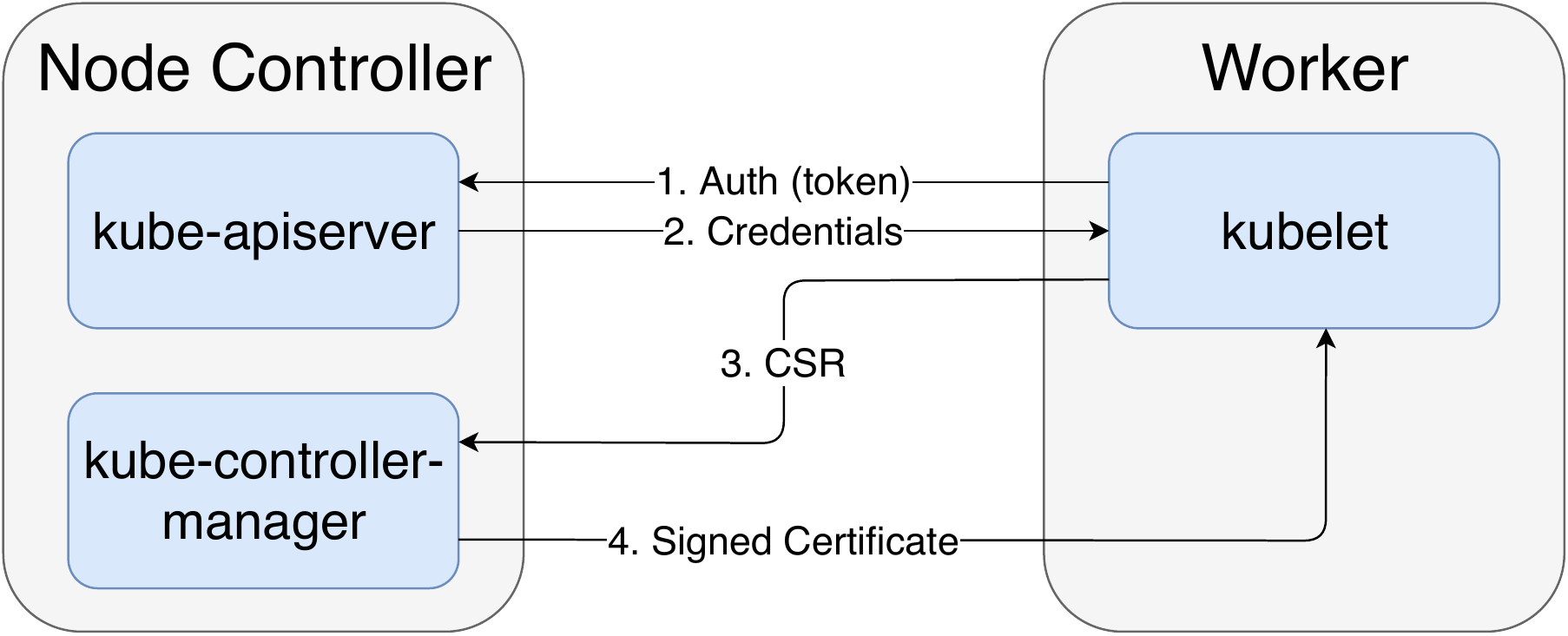}}
 \caption{Kubernetes TLS bootstrap communications. The kubelet uses a limited usage token to authenticate with the API server (Step 1). After authenticating to the API server with the token, the kubelet obtains limited credentials (Step 2) to create a Certificate Signing Request (CSR). The kubelet sends the CSR to the kube-controller-manager (Step 3). The kube-controller-manager can then automatically approve the CSR, or an outside process, possibly a person, can approve the CSR via the Kubernetes API. The kube-controller-manager can then send the created certificate to the kubelet (Step 4), and the kubelet can begin normal operation.}
 \label{fig:kubernetes-tls-bootstrap}
\end{figure}

The first step towards trusting the identity and authentication mechanisms is the bootstrap of the orchestrator (Fig.~\ref{fig:kubernetes-tls-bootstrap}).
In the case of Kubernetes~\cite{kubernetes}, processes called kubelets will start running on the orchestrated machines.
The kubelets are managed by the node controller.
The node controller runs inside the environment of the owner, whereas the kubelets run on all the machines part of the workflow.
First, the kubelet will look for its configuration file containing a key pair and a signed certificate.
The key pair and signed certificate are subsequently used to enable TLS communications from the kubelet to the node controller.
If the kubelet finds it, it can begin normal operation, but if the configuration file does not exist, the kubelet will look for a bootstrap file.

The bootstrap file contains the URL of the node controller, and a limited usage token used to authenticate the kubelet with the node controller (Step 1 of Fig.~\ref{fig:kubernetes-tls-bootstrap}).
After authenticating to the API server with the token, the kubelet obtains limited credentials (Step 2) to create a Certificate Signing Request (CSR).
The kubelet sends the CSR to the kube-controller-manager (Step 3).
The kube-controller-manager can then automatically approve the CSR, or an outside process, possibly a person, can approve the CSR via the Kubernetes API.
Once the CSR is approved, a Certificate is created and issued to the kubelet.
The kube-controller-manager can then send the created certificate to the kubelet (Step 4).
The kubelet can then create a proper configuration file with the key and signed certificate and can begin normal operation.
Optionally, the kubelet can automatically request renewals of the certificate, which goes through the same process as described above.

The orchestrator can now start running the pods on the workers.
Kubernetes will use the previously established secure communication channel with the kubelet to give the kubelet the YAML configuration file of the pods.
The Container Runtime Interface (CRI) on each worker will then make system calls to build the required namespaces until the PID namespace.
At this point, the Istio~\cite{istio} web hook admission controller adds the init containers and the proxy sidecars to the pods, and the Open Policy Agent~\cite{opa} web hook adds the policy sidecars.
The created container images are then started.

\begin{figure}[htbp]
 \centerline{\includegraphics[width=0.55\linewidth]{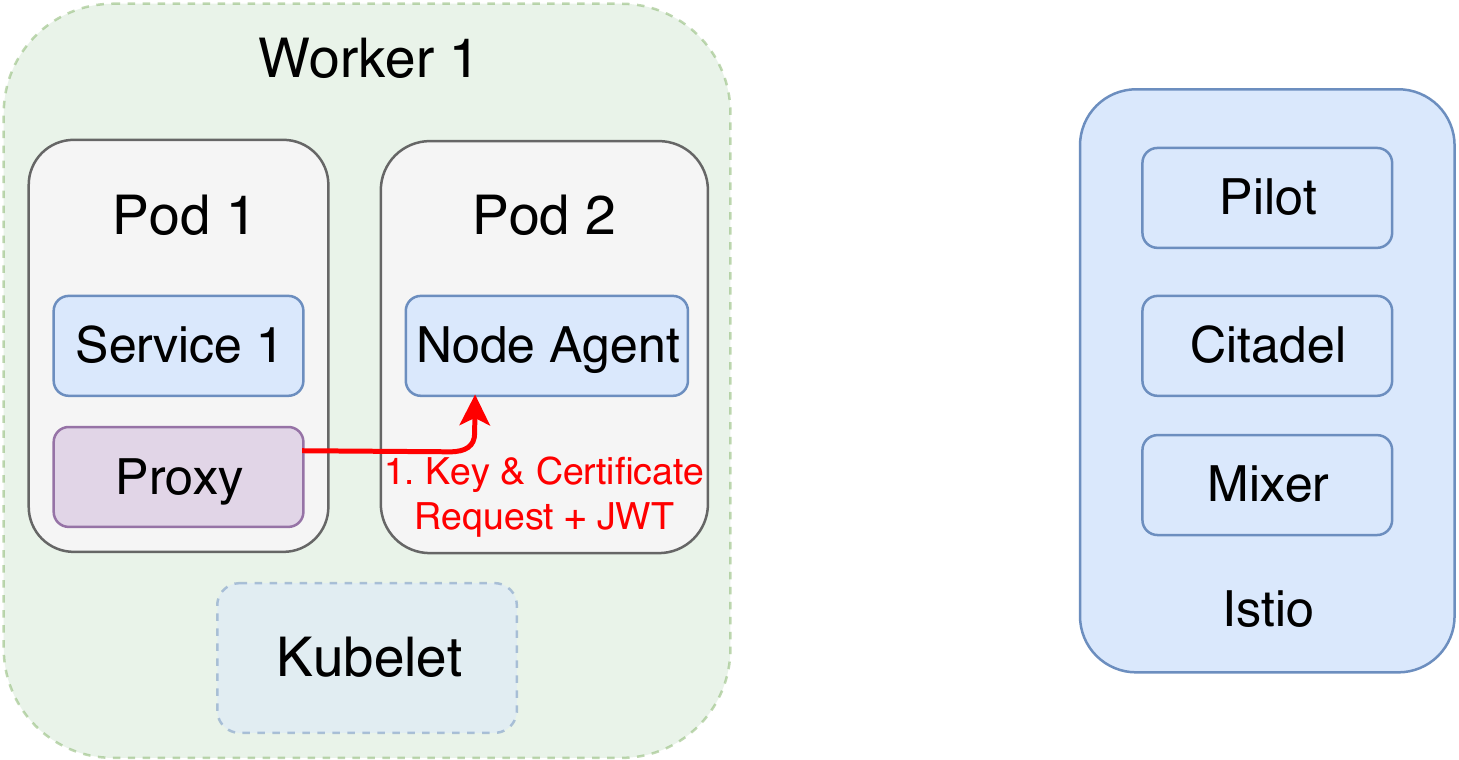}}
 \caption{Istio Key Distribution via Node Agent: Step A. The proxy sends a JWT along an identity request to authenticate it.}
 \label{fig:istio-key-distribution-node-agent-step-1}
\end{figure}

\begin{figure}[htbp]
 \centerline{\includegraphics[width=0.6\linewidth]{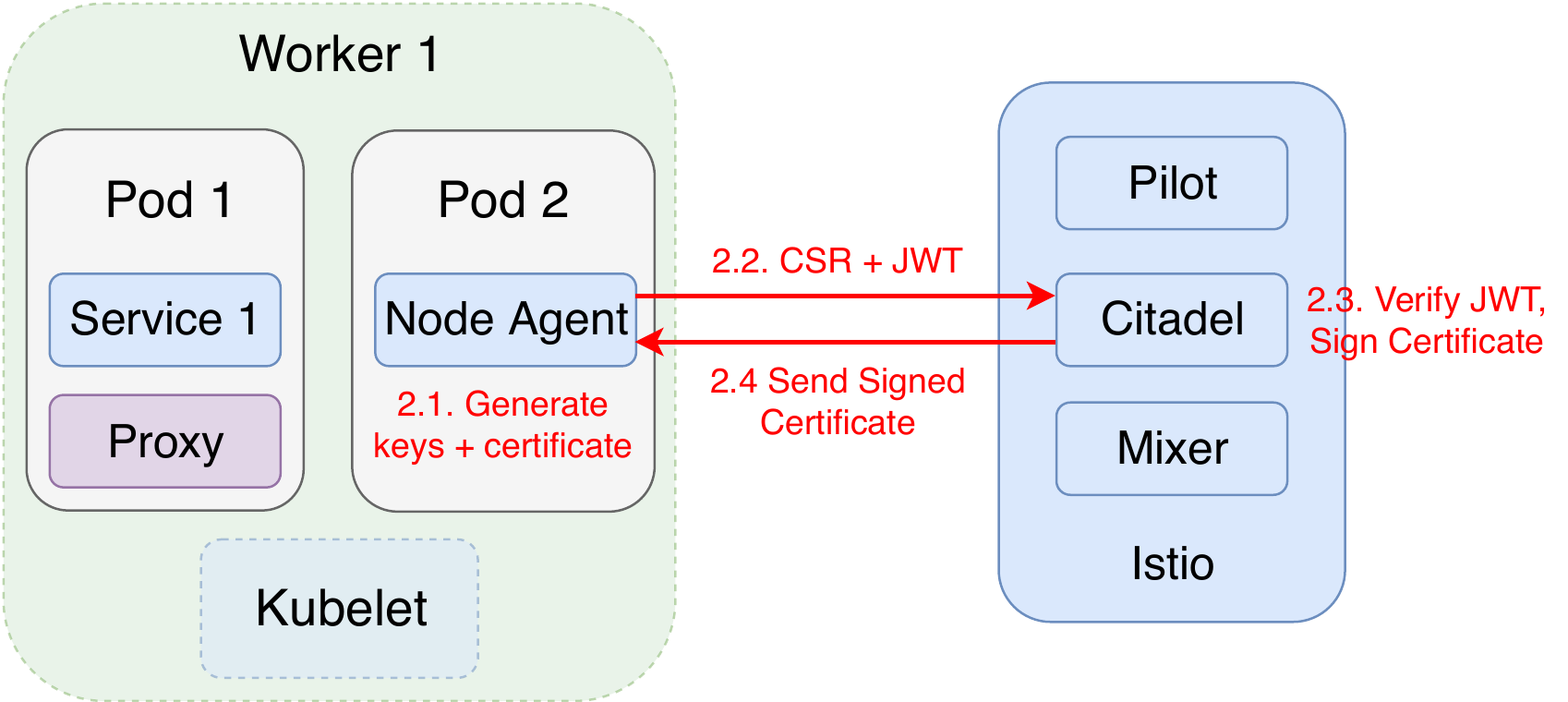}}
 \caption{Istio Key Distribution via Node Agent: Step B. When the Node Agent receives the proxy's request, it generates a key pair and a CSR, and sends the CSR along with the JWT to the Citadel module of Istio. Citadel then authenticates the request and signs the CSR to generate the certificate.}
 \label{fig:istio-key-distribution-node-agent-step-2}
\end{figure}

\begin{figure}[htbp]
 \centerline{\includegraphics[width=0.55\linewidth]{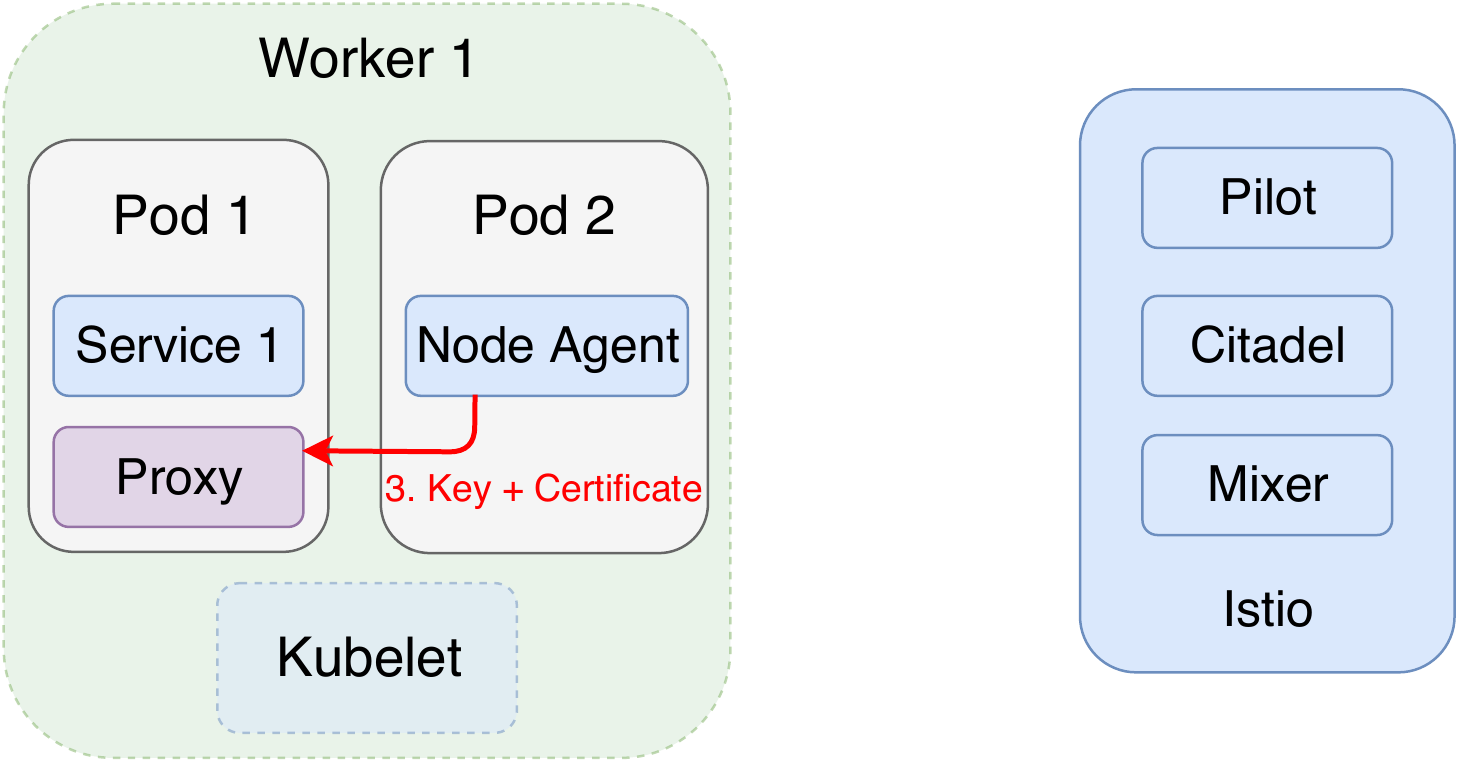}}
 \caption{Istio Key Distribution via Node Agent: Step C. The Node Agent receives the signed certificate and sends it along with the key pair to the proxy that requested the identity.}
 \label{fig:istio-key-distribution-node-agent-step-3}
\end{figure}

Now that kubelets can communicate securely with the node controller and that the pods are running on the orchestrated machines, we need to bootstrap the security of communications between the proxies of our service mesh.
The proxies need to setup their key pair and associated certificate to start communicating in the service mesh via \texttt{mTLS}.
To enable this, a Node Agent running on each worker can receive identity requests from proxies.
The proxy sends a JSON Web Token (JWT) along the request to authenticate it (Fig.~\ref{fig:istio-key-distribution-node-agent-step-1}).
When the Node Agent receives the request, it generates a key pair and a CSR, and sends the CSR along with the JWT to the Citadel module of Istio (Fig.~\ref{fig:istio-key-distribution-node-agent-step-2}).
Citadel then authenticates the request and signs the CSR to generate the certificate.
Once the Node Agent receives the signed certificate, it sends it along with the key pair to the proxy that requested the identity (Fig.~\ref{fig:istio-key-distribution-node-agent-step-3}).
To achieve certificate and key rotation, this process repeats periodically.

Additionally, for the proxies to know which Identity is authorized to run which service, a secure naming information containing this mapping is automatically created from the information retrieved at the Kubernetes node controller.
This is sent by the Pilot module of Istio to each proxy.
With this bootstrap, each proxy can communicate with the others proxies via \texttt{mTLS}, by using their keys to authenticate to each other, and encrypt communications.

\section{Listings}
\label{app:listings}

\begin{lstlisting}[language=json,firstnumber=1, label={lst:ac-ver1}, caption={Formalized Access Control Policy. Each line in role\_perms corresponds to an edge in the workflow depicted in Fig.~\ref{fig:wfs-3}.},captionpos=t]
default allow = false

role_perms = {
  "O":  [{"O": "W",  "dest": "C1_0", "path": "/data"}],
  "C1_0": [{"O": "W",  "dest": "C1_1", "path": "/data"}],
  "C1_0": [{"O": "W",  "dest": "C1_2", "path": "/data"}],
  "C1_1": [{"O": "W",  "dest": "C2", "path": "/data"}],
  "C1_2": [{"O": "W",  "dest": "C4", "path": "/data"}],
  "C2": [{"O": "W",  "dest": "C3",  "path": "/data"}],
  "C3": [{"O": "W",  "dest": "O",  "path": "/data"}],
  "C4": [{"O": "W",  "dest": "O",  "path": "/data"}]
}
\end{lstlisting}

\begin{lstlisting}[language=json,firstnumber=1, label={lst:ac-ver2}, caption={Formalized Access Control Policy with attribute. An attribute is added to specify a time constraint on the communication from $C1_{2}$ to $C4$.},captionpos=t]
default allow = false

role_perms = {
  "O":  [{"O": "W",  "dest": "C1_0", "path": "/data"}],
  "C1_0": [{"O": "W",  "dest": "C1_1", "path": "/data"}],
  "C1_0": [{"O": "W",  "dest": "C1_2", "path": "/data"}],
  "C1_1": [{"O": "W",  "dest": "C2", "path": "/data"}],
  "C1_2": [{"O": "W",  "dest": "C4", "path": "/data"}],
  "C2": [{"O": "W",  "dest": "C3",  "path": "/data"}],
  "C3": [{"O": "W",  "dest": "O",  "path": "/data"}],
  "C4": [{"O": "W",  "dest": "O",  "path": "/data"}]
}

time_const = {
  "C1_2": [{"O": "W",  "dest": "C4", "before": deadline}]
}
\end{lstlisting}

\begin{lstlisting}[language=json,firstnumber=1, label={lst:ac-poc}, caption={Proof of Concept Access Control Policy. This policy implements the permissions described in Table~\ref{tbl:ac}.},captionpos=t]
package istio.authz
import input.attributes.request.http as http_request

default allow = false

# Get username from input
user_name = parsed {
    [_, encoded] := split(http_request.headers.authorization, " ")
    [parsed, _] := split(base64url.decode(encoded), ":")
}

# RBAC user-role assignments
user_roles = {
    "owner": ["owner"],
    "vfx-1": ["vfx-1"],
    "vfx-2": ["vfx-2"],
    "vfx-3": ["vfx-3"],
    "color": ["color"],
    "sound": ["sound"],
    "hdr": ["hdr"]
}

# RBAC role-permissions assignments
role_permissions = {
    "owner": [{"method": "POST",  "path": "/api/vfx-1"}],
    "vfx-1": [{"method": "POST",  "path": "/api/vfx-2"},
              {"method": "POST",  "path": "/api/vfx-3"}],
    "vfx-2": [{"method": "POST",  "path": "/api/color"}],
    "vfx-3": [{"method": "POST",  "path": "/api/sound"}],
    "color": [{"method": "POST",  "path": "/api/hdr"}],
    "hdr": [{"method": "POST",  "path": "/api/owner"}],
    "sound": [{"method": "POST",  "path": "/api/owner"}]
}

# Logic that implements RBAC
rbac_logic {
    # lookup the list of roles for the user
    roles := user_roles[user_name]
    # for each role in that list
    r := roles[_]
    # lookup the permissions list for role r
    permissions := role_permissions[r]
    # for each permission
    p := permissions[_]
    # check if the permission granted to r matches the user's request
    p == {"method": http_request.method, "path": http_request.path}
}


# ABAC user attributes (tenure)
user_attributes = {
    "owner": {"tenure": 8},
    "vfx-1": {"tenure": 3},
    "vfx-2": {"tenure": 12},
    "vfx-3": {"tenure": 7},
    "color": {"tenure": 3},
    "sound": {"tenure": 4},
    "hdr": {"tenure": 5},
}



allow {
  user_name == "owner"

  # Match method and path (RBAC)
  rbac_logic
}

allow {
  user_name == "vfx-1"

  # Match method and path (RBAC)
  rbac_logic
}

allow {
  user_name == "vfx-2"

  # Match method and path (RBAC)
  rbac_logic

  # Match user attributes (ABAC)
  user:=user_attributes[user_name]
  user.tenure > 10
}

allow {
  user_name == "vfx-2"

  # Match method and path (RBAC)
  rbac_logic

  current_time := time.clock([time.now_ns(), "Europe/Paris"])
  to_number(current_time[0]) >= 8
  to_number(current_time[0]) <= 17
}

allow {
  user_name == "vfx-3"

  # Match method and path (RBAC)
  rbac_logic

  # Match user attributes (ABAC)
  user:=user_attributes[user_name]
  user.tenure > 10
}

allow {
  user_name == "vfx-3"

  # Match method and path (RBAC)
  rbac_logic

  current_time := time.clock([time.now_ns(), "Europe/Paris"])
  to_number(current_time[0]) >= 8
  to_number(current_time[0]) <= 17
}

allow {
  user_name == "color"

  # Match method and path (RBAC)
  rbac_logic

  current_time := time.clock([time.now_ns(), "Europe/Paris"])
  to_number(current_time[0]) <= 8
  to_number(current_time[0]) >= 17
}

allow {
  user_name == "sound"

  # Match method and path (RBAC)
  rbac_logic

  current_time := time.clock([time.now_ns(), "Europe/Paris"])
  to_number(current_time[0]) <= 8
  to_number(current_time[0]) >= 17
}

allow {
  user_name == "hdr"

  # Match method and path (RBAC)
  rbac_logic

  current_time := time.clock([time.now_ns(), "Europe/Paris"])
  to_number(current_time[0]) >= 8
  to_number(current_time[0]) <= 17
}
\end{lstlisting}

\end{document}